\documentclass[aps,prl,twocolumn,superscriptaddress,nofootinbib]{revtex4-2}

\usepackage{amsmath,amssymb,amsfonts,bm,mathtools}
\usepackage{hyperref}
\usepackage{tikz}
\usepackage{physics}
\usetikzlibrary{calc,positioning}

\begin{document}

\title{De Sitter Cores from Nonlocal Quantum Field Theories}

\author{E.~J.~Thompson}
\affiliation{Wilfrid Laurier University, Waterloo, ON N2L 3C5, Canada}

\date{\today}

\begin{abstract}
Nonlocal quantum field theories achieve perturbative ultraviolet finiteness by inserting gauge- and diffeomorphism-covariant entire-function regulators into all kinetic terms. These operators can be viewed as nonlocal smearing maps acting on sources in the Einstein equations. In this paper I show that, when the same entire-function regulator responsible for UV-finite loops is applied to the energy--momentum tensor of a point mass, the resulting static, spherically symmetric solution develops a regular de~Sitter core in place of a curvature singularity.
\end{abstract}

\maketitle

\section{Introduction}

In Einstein's general theory of relativity when coupled to reasonable matter sources generically predicts geodesic incompleteness and curvature singularities in gravitational collapse, this indicates the breakdown of the low-energy effective description at short distances~\cite{Raychaudhuri1955,Penrose1965,HawkingEllis1973,Wald1984,WaldGR1984}. One broad approach to ultraviolet (UV) completion is to replace strictly local kinetic operators by covariant entire-function form factors chosen to preserve the infrared limit while softening the ultraviolet behavior of propagators and loop amplitudes~\cite{GE:1967,Moffat1990FiniteNonlocalGauge,EMKW:1991,KW:1992,Krasnikov1987,Tomboulis1997Nonlocal,Moffat2011UVCQG,M:2011a,M:2011b,Modesto2012,Biswas2012GhostFreeGravity,BiswasEtAl2012,ModestoRachwal2017,GM:2021,M:2019,M:2021,LM:2023,BecchiRouetStora1976,Tyutin1975,MT:HUFT-EPJC,MT:Invariant,MT:FiniteHolomorphicQFT,MT:SMmass,MT:SL2C,MT:ReplyToCline,MT:GI2025,MT:AdSdS2025,Thompson2025ToponiumFHQFT,T:DarkMatter2025}.

At the level of static spherically symmetric configurations the effective-source representation of the regulated field equations provides a simple way to connect UV completion to regular black-hole interiors, a number of regular black-hole models replace the $r=0$ singularity by a smooth core, often of de~Sitter type and Gaussian-sourced geometries provide a particularly useful comparison class~\cite{Dymnikova1992,Bardeen1968,Hayward2006,Ansoldi2008,Frolov2015,BuoninfanteKoshelevMazumdar2018,Torres2022,ModestoMoffatNicolini2011,NicoliniSmailagicSpallucci2006}. The point of the present paper is more specific, that once the entire-function regulator is fixed by the UV completion, the same operator that regulates the propagator also determines the source smearing and for the exponential regulator this produces a Gaussian effective energy density and hence a regular de~Sitter core.

The paper is intentionally narrower than a full theory paper as we work in the static Einstein-form effective description, derive the Gaussian profile, solve for the mass function and metric, analyze the induced effective pressures and near-core energy conditions, and then discuss the resulting horizon structure and semiclassical Hawking temperature. The thermodynamic discussion is correspondingly limited, we just use the standard surface-gravity temperature of the static metric and quote the Bekenstein--Hawking area law as the natural effective entropy assignment in the Einstein-form representation~\cite{Bekenstein1973,BardeenCarterHawking1973,Hawking1975,Wald2001,Wald1993Noether,IyerWald1994,MyungKimPark2009,MalufNeves2018}. For the full nonlocal gravitational action the entropy is determined by the Wald--Iyer Noether-charge construction with possible higher-curvature and nonlocal corrections~\cite{Wald1993,IyerWald1994,JacobsonKangMyers1994,Visser1993Surface,ConroyMazumdarTeimouri2015,Myung2017}. We do not attempt in this work a full Wald derivation from the complete nonlocal action, nor do we address dynamical collapse, time-dependent smoothing, backreaction, or the stability of the inner horizon or of any remnant endpoint, that is saved for a follow up paper.

\section{Nonlocal field equations and the effective-source representation}

To start we let $g_{\mu\nu}$ be a Lorentzian metric with signature $(-,+,+,+)$ and let $D_\mu$ denote the appropriate gauge and diffeomorphism covariant derivative. The nonlocal regulator is taken to be an entire function of the covariant d'Alembertian:
\begin{equation}
\Box \equiv -g^{\mu\nu}D_\mu D_\nu.
\end{equation}
We write the regulated gravitational field equations as:
\begin{equation}
F^{-2}\!\left(\frac{\Box}{\Lambda_G^2}\right)G_{\mu\nu}=\kappa T_{\mu\nu},
\qquad \kappa\equiv 8\pi G_N,
\label{eq:geom_form}
\end{equation}
where $\Lambda_G$ is the gravitational nonlocality scale. The only properties of $F$ needed in what follows are that it is entire, nonvanishing in the finite complex plane, and normalized so that $F(0)=1$. These assumptions guarantee that the operator is invertible on the relevant function space and that no additional poles are introduced into the propagator~\cite{Moffat1990FiniteNonlocalGauge,EMKW:1991,KW:1992,Krasnikov1987,Tomboulis1997Nonlocal,Biswas2012GhostFreeGravity,BiswasEtAl2012,Modesto2012,ModestoRachwal2017,MT:GI2025}.

Equation~\eqref{eq:geom_form} can then be rewritten in Einstein form:
\begin{equation}
G_{\mu\nu}=\kappa S_{\mu\nu},
\qquad
S_{\mu\nu}\equiv F^2\!\left(\frac{\Box}{\Lambda_G^2}\right)T_{\mu\nu},
\label{eq:einstein_form}
\end{equation}
within the semiclassical linear response regime relevant here this is not an additional physical postulate but an equivalent representation of the same regulated dynamics. Linearizing about Minkowski space, $g_{\mu\nu}=\eta_{\mu\nu}+\kappa h_{\mu\nu}$ in harmonic gauge we find:
\begin{equation}
F^{-2}\!\left(\frac{\Box}{\Lambda_G^2}\right)\Box \bar h_{\mu\nu}=-2\kappa T_{\mu\nu},
\end{equation}
so multiplication by $F^2$ yields:
\begin{equation}
\Box \bar h_{\mu\nu}=-2\kappa S_{\mu\nu}.
\end{equation}
The same entire-function form factor that regulates the propagator therefore determines the dressed source. In what follows we use the Einstein-form representation~\eqref{eq:einstein_form}, because it isolates the effect of the regulator on the gravitating source while preserving the usual geometric interpretation of the metric.

For the static configurations of interest we specialize to the exponential regulator in its effective-source form:
\begin{equation}
F^2\!\left(\frac{\Box}{\Lambda_G^2}\right)\Big|_{\rm static}
=
\exp\!\left(\frac{\nabla^2}{\Lambda_G^2}\right),
\label{eq:static_reg}
\end{equation}
where $\nabla^2$ is the flat spatial Laplacian in the weak-field short-distance regime. With these sign conventions the regulator acts as a heat-kernel smearing operator producing Gaussian suppression in momentum space and Gaussian smoothing in position space~\cite{Vassilevich2003,NicoliniSmailagicSpallucci2006}.

\section{Static point source and Gaussian effective density}

I we consider a static point mass $M$ at the origin the bare rest-frame energy density is:
\begin{equation}
\rho_{\rm bare}(\mathbf{x})=M\,\delta^{(3)}(\mathbf{x}),
\end{equation}
with bare stress tensor:
\begin{equation}
T^0{}_0(\mathbf{x})=-\rho_{\rm bare}(\mathbf{x}),
\qquad
T^i{}_j(\mathbf{x})=0,
\end{equation}
we see that the effective energy density that gravitates is then:
\begin{equation}
\rho_{\rm eff}(\mathbf{x})\equiv -S^0{}_0(\mathbf{x})
=
\exp\!\left(\frac{\nabla^2}{\Lambda_G^2}\right)\rho_{\rm bare}(\mathbf{x}).
\end{equation}
Using the Fourier representation:
\begin{equation}
\delta^{(3)}(\mathbf{x})=\int \frac{d^3k}{(2\pi)^3}e^{i\mathbf{k}\cdot \mathbf{x}},
\qquad
\nabla^2 e^{i\mathbf{k}\cdot\mathbf{x}}=-k^2 e^{i\mathbf{k}\cdot\mathbf{x}},
\end{equation}
we find:
\begin{align}
\rho_{\rm eff}(\mathbf{x})
&=
M\int\frac{d^3k}{(2\pi)^3}
\exp\!\left(-\frac{k^2}{\Lambda_G^2}\right)e^{i\mathbf{k}\cdot\mathbf{x}}
\nonumber\\
&=
\frac{M\Lambda_G^3}{(4\pi)^{3/2}}
\exp\!\left(-\frac{\Lambda_G^2 r^2}{4}\right),
\qquad r\equiv |\mathbf{x}|.
\label{eq:rhoeff}
\end{align}
So the point source is replaced by a smooth Gaussian ball of width $\ell_{\rm nl}\sim \Lambda_G^{-1}$, and the total mass is preserved:
\begin{equation}
\int d^3x\,\rho_{\rm eff}(\mathbf{x})=M.
\end{equation}
The core density is finite:
\begin{equation}
\rho_{\rm eff}(0)=\frac{M\Lambda_G^3}{(4\pi)^{3/2}}.
\end{equation}

\section{Static spherically symmetric geometry and the de~Sitter core}

I we now assume a static spherically symmetric metric in Schwarzschild gauge:
\begin{equation}
ds^2=-f(r)dt^2+\frac{dr^2}{f(r)}+r^2d\Omega^2,
\label{eq:sss_metric}
\end{equation}
then introduce the Misner--Sharp mass function~\cite{MisnerSharp1964},
\begin{equation}
f(r)=1-\frac{2G_N m(r)}{r}.
\label{eq:f_m}
\end{equation}
For a static spherically symmetric source $S^\mu{}_{\nu}=\mathrm{diag}(-\rho_{\rm eff},p_r,p_\perp,p_\perp)$, the $tt$ Einstein equation gives the standard relation:
\begin{equation}
m'(r)=4\pi r^2\rho_{\rm eff}(r).
\label{eq:mprime}
\end{equation}
Substituting Eq.~\eqref{eq:rhoeff} and integrating from $0$ to $r$ gives us:
\begin{equation}
m(r)=M\left[\operatorname{erf}\!\left(\frac{\Lambda_G r}{2}\right)-\frac{\Lambda_G r}{\sqrt{\pi}}e^{-\Lambda_G^2 r^2/4}\right]
=
M\,\frac{\gamma\!\left(\frac32;\frac{\Lambda_G^2 r^2}{4}\right)}{\Gamma\!\left(\frac32\right)}.
\label{eq:massfunction}
\end{equation}
As $r\to\infty$, $m(r)\to M$ and we recover the ordinary Schwarzschild exterior~\cite{Schwarzschild1916}. For $r\Lambda_G\ll 1$ the Gaussian density is approximately constant:
\begin{equation}
\rho_{\rm eff}(r)=\rho_0+\mathcal O(r^2),
\qquad
\rho_0\equiv \frac{M\Lambda_G^3}{(4\pi)^{3/2}},
\end{equation}
so Eq.~\eqref{eq:mprime} gives:
\begin{equation}
m(r)=\frac{4\pi}{3}\rho_0 r^3+\mathcal O(r^5)
=
\frac{M\Lambda_G^3}{6\sqrt{\pi}}r^3+\mathcal O(r^5).
\end{equation}
Then Eq.~\eqref{eq:f_m} becomes:
\begin{equation}
f(r)=1-\frac{G_N M\Lambda_G^3}{3\sqrt{\pi}}r^2+\mathcal O(r^4)
=
1-\frac{\Lambda_{\rm eff}}{3}r^2+\mathcal O(r^4),
\label{eq:f_smallr}
\end{equation}
with
\begin{equation}
\Lambda_{\rm eff}=\frac{G_N M\Lambda_G^3}{\sqrt{\pi}}.
\label{eq:Lambdaeff}
\end{equation}
The core geometry is therefore the static patch of de~Sitter space:
\begin{equation}
ds^2\approx -\left(1-\frac{\Lambda_{\rm eff}}{3}r^2\right)dt^2
+\frac{dr^2}{1-\frac{\Lambda_{\rm eff}}{3}r^2}+r^2 d\Omega^2,
\qquad r\Lambda_G\ll 1.
\end{equation}
In particular:
\begin{equation}
R(0)=4\Lambda_{\rm eff}=\frac{4G_N M\Lambda_G^3}{\sqrt{\pi}},
\end{equation}
and all curvature invariants remain finite at the center.

This de~Sitter behavior is not peculiar to the Gaussian profile alone. More generally, any smooth effective density with finite positive central value $\rho_{\rm eff}(0)=\rho_0$ gives $m(r)=\frac{4\pi}{3}\rho_0 r^3+\mathcal O(r^5)$ and hence $f(r)=1-\frac{8\pi G_N\rho_0}{3}r^2+\mathcal O(r^4)$. The role of the exponential regulator is therefore to provide an explicit, ultraviolet-motivated realization of a mechanism whose small-$r$ consequence is generic for regular spherically symmetric cores~\cite{Dymnikova1992,Bardeen1968,Hayward2006,Ansoldi2008,Frolov2015,BuoninfanteKoshelevMazumdar2018,Torres2022}. What the regulator fixes in the present model is not merely the existence of the core but its detailed profile and the relation of its curvature scale to the ultraviolet scale $\Lambda_G$.

Within the effective description used here freely falling minimally coupled test bodies probe the geometry determined by Eq.~\eqref{eq:sss_metric}. In that sense the de~Sitter core has the usual metric meaning, that in the small-$r$ region where the static effective description is valid, test particles follow geodesics of the regular metric. What is altered by the nonlocality is the source that appears on the right-hand side of Einstein's equations, not the kinematics of test bodies once the effective metric has been obtained.

\section{Effective pressures and energy conditions}

In Schwarzschild gauge we have $G^t{}_t=G^r{}_r$ so then Einstein's equations imply:
\begin{equation}
p_r(r)=-\rho_{\rm eff}(r).
\label{eq:pr}
\end{equation}
Conservation of the effective stress tensor, $\nabla_\mu S^{\mu\nu}=0$, then gives us:
\begin{equation}
p_\perp(r)=p_r(r)+\frac{r}{2}p_r'(r)
=
-\rho_{\rm eff}(r)-\frac{r}{2}\rho_{\rm eff}'(r),
\label{eq:pperp_general}
\end{equation}
for the Gaussian profile:
\begin{equation}
\rho_{\rm eff}'(r)=-\frac{\Lambda_G^2 r}{2}\rho_{\rm eff}(r),
\end{equation}
and therefore:
\begin{equation}
p_\perp(r)=\rho_{\rm eff}(r)\left(\frac{\Lambda_G^2 r^2}{4}-1\right).
\label{eq:pperp_gaussian}
\end{equation}
Near the origin of the black hole:
\begin{equation}
p_r(0)\simeq p_\perp(0)\simeq -\rho_0,
\end{equation}
so the effective equation of state becomes de~Sitter-like.

The strong energy condition combination is:
\begin{equation}
\rho_{\rm eff}+p_r+2p_\perp=2p_\perp
=
2\rho_{\rm eff}(r)\left(\frac{\Lambda_G^2 r^2}{4}-1\right),
\end{equation}
hence the strong energy condition is violated for:
\begin{equation}
r<\frac{2}{\Lambda_G}\sim \mathcal O(\ell_{\rm nl}),
\end{equation}
and recovered outside that scale. This behavior is expected as the same nonlocal smearing that resolves the singularity also produces the effective negative pressure required to evade the usual focusing argument behind the classical singularity theorems.

\section{Horizon structure and semiclassical temperature}

Horizons occur at positive roots of:
\begin{equation}
f(r_h)=0
\qquad\Longleftrightarrow\qquad
r_h=2G_N m(r_h),
\label{eq:horizon}
\end{equation}
because $m(r)\sim r^3$ for small $r$ and $m(r)\to M$ for large $r$ Eq.~\eqref{eq:horizon} can admit two horizons, one degenerate horizon, or no horizon, depending on the dimensionless combination $G_N M\Lambda_G$. The extremal case additionally satisfies:
\begin{equation}
f'(r_0)=0
\qquad\Longleftrightarrow\qquad
m'(r_0)=\frac{m(r_0)}{r_0}.
\label{eq:extremalcondition}
\end{equation}
Using Eq.~\eqref{eq:mprime} together with Eq.~\eqref{eq:horizon} this becomes:
\begin{equation}
8\pi G_N r_0^2\rho_{\rm eff}(r_0)=1.
\label{eq:extremal_density}
\end{equation}
For the static metric~\eqref{eq:sss_metric}, the surface gravity at a simple
Killing horizon is $\kappa_h=\tfrac12 f'(r_h)$, so the semiclassical Hawking temperature is~\cite{Bekenstein1973,BardeenCarterHawking1973,Hawking1975,GibbonsHawking1977,Wald2001}:
\begin{equation}
T_H(r_h)=\frac{f'(r_h)}{4\pi}.
\label{eq:TH_def}
\end{equation}
In the Einstein-form effective-source description this coincides with the usual Hawking temperature. In the full nonlocal gravitational theory the thermodynamic temperature must be obtained from the Noether-charge entropy by
\(T_{\rm th}^{-1}=\partial S_{\rm Wald}/\partial M\). From Eq.~\eqref{eq:f_m} we obtain:
\begin{equation}
f'(r)=2G_N\left(\frac{m(r)}{r^2}-\frac{m'(r)}{r}\right),
\end{equation}
and therefore, at the horizon:
\begin{equation}
T_H(r_h)=\frac{1}{4\pi r_h}\left[1-2G_N m'(r_h)\right]
=
\frac{1}{4\pi r_h}\left[1-8\pi G_N r_h^2\rho_{\rm eff}(r_h)\right].
\label{eq:TH_general}
\end{equation}
For \(r_h\gg \Lambda_G^{-1}\) the effective density at the horizon
is exponentially small and the surface-gravity expression reduces
to the Schwarzschild value:
\begin{equation}
T_{\rm sg}\simeq \frac{1}{4\pi r_h}
=\frac{1}{8\pi G_NM}.
\end{equation}
This exponential suppression applies only to the effective-source
contribution appearing explicitly in \eqref{eq:TH_general}. It should not be confused with the Wald-entropy correction of the full nonlocal gravitational action that is controlled by curvature invariants at the horizon and is therefore generically power-suppressed in \((\Lambda_G r_h)^{-1}\), not exponentially suppressed. When $r_h\gg \Lambda_G^{-1}$ the density at the horizon is exponentially small and Eq.~\eqref{eq:TH_general} reduces to the Schwarzschild result:
\begin{equation}
T_H\simeq \frac{1}{4\pi r_h}=\frac{1}{8\pi G_N M}.
\end{equation}
At extremality, Eq.~\eqref{eq:extremal_density} implies for us:
\begin{equation}
T_H(r_0)=0,
\end{equation}
so the static semiclassical solution is compatible with a zero-temperature finite-mass endpoint. The sign of the near-extremal heat capacity, the dynamical stability of this endpoint, and the role of evaporation backreaction require a separate analysis and are not part of the present paper but is saved for future work.

For the Gaussian profile~\eqref{eq:rhoeff} it is useful to define $x_h\equiv \Lambda_G r_h/2$ and:
\begin{equation}
\mathcal F(x)\equiv \operatorname{erf}(x)-\frac{2x}{\sqrt{\pi}}e^{-x^2},
\end{equation}
so that $m(r)=M\mathcal F(\Lambda_G r/2)$. Then the horizon relation becomes:
\begin{equation}
M(r_h)=\frac{r_h}{2G_N\mathcal F(x_h)},
\end{equation}
while the temperature can be written as:
\begin{equation}
T_H(r_h)=\frac{1}{4\pi r_h}
\left[
1-\frac{4x_h^3 e^{-x_h^2}}{\sqrt{\pi}\,\mathcal F(x_h)}
\right].
\end{equation}
The extremality condition is therefore:
\begin{equation}
1=\frac{4x_0^3 e^{-x_0^2}}{\sqrt{\pi}\,\mathcal F(x_0)}.
\end{equation}
Numerically we find $x_0\simeq 1.5112$, and then we see:
\begin{equation}
r_0\simeq \frac{3.0224}{\Lambda_G},
\qquad
M_0\simeq \frac{1.904}{G_N\Lambda_G}.
\end{equation}

Because the field equations are written in Einstein form with constant Newton coupling the natural semiclassical entropy assignment for the outer horizon is the usual Bekenstein--Hawking area law~\cite{Wald1993,IyerWald1994,JacobsonKangMyers1994,Visser1993Surface}:
\begin{equation}
S_{\rm BH}(r_+)=\frac{A_+}{4G_N}=\frac{\pi r_+^2}{G_N}.
\end{equation}
In the present paper this should be understood as an effective entropy assignment appropriate to the Einstein-form representation of the static solution. A first-principles Wald entropy for the underlying nonlocal action would require the full variational treatment of the nonlocal operator and lies beyond the scope of the present work~\cite{Wald2001,Wald1993Noether,IyerWald1994}. For related thermodynamic analyses of regular black holes, see~\cite{Ansoldi2008,MyungKimPark2009,MalufNeves2018,Torres2022}.

The entropy formula should not be interpreted as the exact entropy of the full nonlocal gravitational action, it is the natural area assignment in the Einstein-form effective-source representation where the geometric field equations are written as \(G_{\mu\nu}=\kappa S_{\mu\nu}\) with constant \(G_N\). If instead we vary the underlying nonlocal gravitational action directly the black-hole entropy is determined by the Wald--Iyer Noether charge rather than by the area term alone.

To see the size and structure of the correction we can consider the local derivative expansion of the nonlocal gravitational action:
\begin{widetext}
\begin{equation}
\begin{aligned}
\begin{split}
I_{\rm grav}
=
\frac{1}{16\pi G_N}
\int d^4x\sqrt{-g}
\left[
R
+
\frac{1}{\Lambda_G^2}
\left(
\alpha R^2
+\beta R_{\mu\nu}R^{\mu\nu}
+\gamma R_{\mu\nu\rho\sigma}R^{\mu\nu\rho\sigma}
\right)
+O(\Lambda_G^{-4})
\right],
\end{split}
\end{aligned}
\end{equation}
\end{widetext}
where the dimensionless coefficients \(\alpha,\beta,\gamma\) are determined by the small-\(\Box/\Lambda_G^2\) expansion of the chosen nonlocal form factor. For any stationary black hole with bifurcate Killing horizon \(\mathcal H\), binormal \(\epsilon_{ab}\), induced metric \(h_{ij}\), and Lagrangian density \(\mathcal L\) the entropy is:
\begin{equation}
S_{\rm Wald}
=
-2\pi
\int_{\mathcal H}
d^2x\sqrt{h}\,
\frac{\partial \mathcal L}{\partial R_{abcd}}
\epsilon_{ab}\epsilon_{cd},
\end{equation}
so we see:
\begin{equation}
S_{\rm Wald}
=
\frac{A_h}{4G_N}
+
\Delta S_{\rm Wald},
\end{equation}
where:
\begin{widetext}
\begin{equation}
\begin{aligned}
\begin{split}
\Delta S_{\rm Wald}
=
-2\pi
\int_{\mathcal H}
d^2x\sqrt{h}\,
\frac{1}{16\pi G_N\Lambda_G^2}
\Bigg[
\alpha R
\left(g^{ac}g^{bd}-g^{ad}g^{bc}\right)
+
\frac{\beta}{2}
\left(
g^{ac}R^{bd}
-g^{ad}R^{bc}
-g^{bc}R^{ad}
+g^{bd}R^{ac}
\right)
+
2\gamma R^{abcd}
\Bigg]
\epsilon_{ab}\epsilon_{cd},
\end{split}
\end{aligned}
\end{equation}
\end{widetext}
and we add a $O(\Lambda_G^{-4})$ contribution. So correction is controlled by the curvature at the horizon, and since \(\mathcal R_h\sim r_h^{-2}\) for a large black hole,the correction scales as:
\begin{equation}
\frac{\Delta S_{\rm Wald}}{S_{\rm BH}}
=
O\!\left(\frac{1}{\Lambda_G^2 r_h^2}\right),
\end{equation}
up to coefficients fixed by the detailed gravitational form factor. This contribution is power-suppressed in the infrared but it is not exponentially suppressed by the Gaussian tail of \(\rho_{\rm eff}(r_h)\). The reason is that the Noether charge depends on the curvature of the horizon geometry itself while including the Schwarzschild-like Weyl curvature and not only on the local value of the effective matter density.

We see that the surface-gravity expression:
\begin{equation}
T_{\rm sg}
=
\frac{\kappa_h}{2\pi}
=
\frac{f'(r_h)}{4\pi},
\end{equation}
should be understood as the geometric Hawking temperature associated with the effective static metric. The thermodynamic temperature of the full nonlocal theory is instead determined by:
\begin{equation}
T_{\rm th}^{-1}
=
\frac{\partial S_{\rm Wald}}{\partial M},
\end{equation}
or if we are using \(r_h\) as a parameter, then:
\begin{equation}
T_{\rm th}(r_h)
=
\left(
\frac{dS_{\rm Wald}/dr_h}{dM/dr_h}
\right)^{-1}.
\end{equation}
So:
\begin{equation}
T_{\rm th}
=
T_{\rm sg}
\left[
1
+
O\!\left(\frac{1}{\Lambda_G^2 r_h^2}\right)
\right]^{-1},
\end{equation}
where the precise coefficient is model-dependent. In this paper we only use \(T_{\rm sg}\) as an effective surface-gravity diagnostic of the static metric and do not claim that the area law gives the exact entropy of the full nonlocal action.

So our nonlocal regulator modifies the entropy through the Noether charge at power-law order while the effective source density at the horizon may be exponentially small. There are related analyses of black-hole entropy and horizon structure in infinite-derivative and weakly nonlocal gravity~\cite{ConroyMazumdarTeimouri2015,ConroyMazumdarTalaganisTeimouri2015,Myung2017,BuoninfanteGiacchiniNetto2024}.

\section{Remarks on the maximal analytic extension}

For completeness of this paper it is useful to record the causal interpretation of the exact static solution as the maximal analytic extension of a static black-hole metric is obtained by the standard Eddington--Finkelstein and Kruskal constructions~\cite{Finkelstein1958,Kruskal1960,ReallNotesBH2014,Wald1984,WaldGR1984,GaurVisser2024BHWH}. In such an eternal extension a white-hole region appears as the time reverse of the black-hole region. This statement is kinematical and follows from the extension of the static metric itself but it is not a claim that realistic astrophysical collapse produces a white hole.

In the present class of geometries the small-$r$ singular boundary is replaced by a regular core. Accordingly the Penrose diagram is Reissner--Nordstr\"om-like in the two-horizon regime, except that the $r=0$ singularity is replaced by a regular timelike center. At the same time we must distinguish carefully between this idealized eternal extension and realistic collapse. Classical white holes are unstable to perturbations~\cite{Eardley1974DeathWH,BarrabesIsraelPoisson1993DeathWH,Nikitin2018StabilityWH,BarceloEtAl2015TransientWH}, and inner Cauchy horizons are well known to be sensitive to mass inflation~\cite{PoissonIsrael1990,Ori1991,PoissonIsrael1990MassInflation,Ori1991MassInflation,ConroyMazumdarTeimouri2015,ConroyKoshelevMazumdar2015,Myung2017,BuoninfanteGiacchiniNetto2024}. The presence of a regular core therefore does not by itself establish dynamical stability of the full interior. Those questions lie beyond the static analysis given here.

For similar reasons the existence of a zero-temperature extremal endpoint in the present semiclassical treatment should not be conflated with specific black-to-white-hole or remnant scenarios in the quantum-gravity literature~\cite{RovelliVidotto2024PlanckStarsWHReview,BianchiEtAl2018WHRemnants,BarrauEtAl2021CloserLookWHRemnants,HanRovelliSoltani2023GeometryBHtoWH,Ong2024RemnantsReview}. The result established here is more modest, that within the static effective-source description generated by the entire-function regulator, the horizon temperature can vanish at finite mass.

\section{Conclusion}

The main result of this paper is straightforward as in the static Einstein-form representation of an entire-function regulated nonlocal gravity theory a point mass is mapped to the Gaussian effective density~\eqref{eq:rhoeff}. Solving the corresponding spherically symmetric Einstein equations yields the exact mass function~\eqref{eq:massfunction}, an asymptotically Schwarzschild exterior, and a de~Sitter core at small radius with effective cosmological constant~\eqref{eq:Lambdaeff}. The same solution induces a de~Sitter-like equation of state near the center, confines strong-energy-condition violation to the nonlocal core scale, admits an outer horizon, inner horizon, or extremal configuration depending on $G_N M\Lambda_G$, and gives the semiclassical temperature~\eqref{eq:TH_general}, which vanishes at extremality.

Just as important are the limits of the result, the paper does not claim to solve the full nonlocal initial-value problem as to prove that arbitrary time-dependent collapse is smoothed in the same way or to establish the dynamical stability of the inner horizon or of the near-extremal endpoint. It also does not provide a full Wald derivation of the entropy from the nonlocal action. What it does show is that the same ultraviolet regulator used to render the quantum theory finite naturally produces, in the static effective-source description, a nonsingular de~Sitter core in place of the classical Schwarzschild singularity.

\section*{Acknowledgements}
I would like to thank Hilary Carteret, John Moffat, and Niayesh Afshordi for helpful discussions. I would as well like to thank the refree for very helpful comments.

\section*{Data availability}
This manuscript is a theoretical study. No new data were generated or analyzed in the course of this work.

\bibliographystyle{unsrt}

\end{document}